\documentstyle[epsfig,aps,amsmath]{revtex}

\begin{document}

\draft

\title{Kinetic energy of a trapped Fermi gas interacting with a
  Bose-Einstein condensate}

\author{L. Vichi$^a$, M. Amoruso$^b$, A. Minguzzi$^b$, S.
Stringari$^a$ and M. P. Tosi$^b$} 
\address{$^a$Istituto Nazionale per la Fisica della Materia and Dipartimento di
Fisica, Universit\`a di Trento, Via Sommarive 14, I-38050 Povo, Italy}
\address{$^b$Istituto Nazionale per la Fisica della Materia and Classe di
Scienze, Scuola Normale Superiore, Piazza dei Cavalieri
7, I-56126  Pisa, Italy}

\date{September 10, 1999}
\maketitle

\begin{abstract}

We study a confined mixture of bosons and fermions in the regime of
quantal degeneracy, with particular attention to the effects of the
interactions on the kinetic energy of the fermionic component.
We are able to explore a wide region of system parameters by identifying two 
scaling variables which completely determine its state  
at low temperature. These are the ratio of the boson-fermion and
boson-boson interaction strengths and the ratio of the radii of the two
clouds.
We find that the effect of the interactions can be sizeable for
reasonable choices of the parameters and that its experimental study
can be used to infer the sign of
the boson-fermion scattering length.
The interplay between
interactions and thermal effects in the fermionic kinetic
energy is also discussed.
\end{abstract}

\pacs{03.75.Fi, 67.40.Db, 67.40.Kh, 67.60.-g}

\section{Introduction}
\label{sec:intro}

The  achievement of Bose-Einstein condensation in trapped gases
\cite{BEC,BEC2,BEC3} has opened new opportunities
for investigating the low temperature behaviour of
dilute quantum  systems. 
Recent experimental studies have been also addressed to
the search of degeneracy effects in Fermi gases
\cite{Jin,Prevedelli}
and in mixtures of bosons and fermions
\cite{preprintparigi}. First experimental evidences of these effects
have been recently reported in \cite{Demarco}.
Due to the Pauli 
exclusion principle the effects of the interactions in a Fermi gas are
much weaker than for a Bose condensate
\cite{Silvera,Butts,Schneider,Burnett,Stoof,Molmer}.
Therefore, the kinetic energy dominates the behaviour of the Fermi gas
and is a clear
indicator of its 
quantum degeneracy. 

Interest in boson-fermion mixtures is stimulated by the fact that,
whereas in a one-component spin-polarized Fermi
gas the absence of interactions leads to long thermalization times
which hamper the process of evaporative cooling,
the collisions between the two
species in a mixture can ensure fast thermalization (the so-colled
sympatethic cooling \cite{Myatt,Timmermans,Geist}).
The kinetic energy of the Fermi
component in such a mixture could be measured by time-of-flight 
techniques in an ideal experiment in which the confining potential is
suddenly  switched off after a fast expulsion (\emph{i.e.}~on a time scale
shorter than the boson-fermion collision time) of the bosons from the
trap. In this way one can
avoid 
the effects of the interactions during the expansion of the gas.
Alternatively, the
kinetic energy could be obtained from inelastic photon scattering at
high momentum transfer as recently shown in the case
of a trapped Bose gas \cite{bragg}.

In this paper we analyze the behaviour of the kinetic energy
of the fermionic component in a boson-fermion mixture.
We find that the kinetic energy can be significantly affected by the
interactions of the Fermi gas with the Bose-Einstein condensed cloud
for reasonable choices of the system parameters.
An important consequence of this purely quantum
effect is that one can measure the sign and the strength
of the boson-fermion
scattering length by using the Bose
component as a tunable device to change the effective
potential felt by the fermions \cite{TnFi}.
We assume a positive boson-boson scattering
length, with a view to applications to mixtures of Rb-K and Na-K.

\section{Interacting Fermi-Bose mixtures}
\label{sec1}

For the description of the mixture at finite temperature we adopt  
the semiclassical three-fluid
model already developed in ref.\cite{PiTn1}. We consider a system of  $N_f$ 
fermions of mass
$m_f$ and $N_b$ bosons of mass $m_b$ confined by  external potentials
$  V_{ext}^{f,b}(r)= \frac 1 2 m_{f,b} \omega^2_{f,b} r^2$
with frequencies $\omega_f$ and $\omega_b$.
The external potentials are assumed to be spherically symmetric, the
asymmetric case requiring simply a change of variables in the framework of
the semiclassical approximation that we adopt.
We include the interaction between bosons through the scattering
length $a_{bb}$ and the interaction between bosons and fermions
through the scattering length $a_{bf}$.
By assuming that a single spin state is trapped for each component
of the mixture, we can safely neglect the fermion-fermion
interaction which is inhibited by the Pauli exclusion principle. 
Extensions to multi-spin configurations can be naturally
made within the present formalism. In the following we will neglect
the possibility of a superfluid phase for the fermionic component (for
a discussion of the BCS transition in trapped Fermi gases see
\cite{Kagan,Stoof2}) as well as the possibility of the expulsion of the bosons
from the center of the cloud 
(this phase separation is expected to occur
only for values of the coupling strengths such that the
mean field contribution of the fermions is comparable with the mean
field contribution of the bosons. This requires values of the
parameters such that $n_fa_{bb}^3 \sim
1$. The resulting system is no more a dilute one, see
\cite{Molmer2,Viverit}).

The spatial densities of the  condensed bosons ($n_c$),
of the bosonic thermal component ($n_{nc}$) and of the
fermions ($n_f$) are determined by the self-consistent solution of the
following equations:
\begin{align} \label{nc}
  n_c(r)&= \frac 1 {g_{bb}} \left(\mu_b -V_{ext}^b(r)-2g_{bb}n_{nc}(r)
    -g_{bf}n_f(r) \right),\\ 
  \label{nnc}
  n_{nc}(r)&= \int \frac{d^3p}{(2 \pi \hbar)^3} \left( \exp [\beta
    (\frac{p^2}{2m_b}
    +V_{eff}^b(r)-\mu_b)]-1 \right)^{-1}  
\end{align}
{\rm and}  
\begin{align}
\label{nf}
  n_f(r)&= \int \frac{d^3p}{(2 \pi \hbar)^3} \left( \exp [ \beta
    (\frac{p^2}{2m_f}
    +V_{eff}^f(r)-\mu_f)]+1 \right)^{-1}.
\end{align}
Here, the effective potentials acting on the thermal boson cloud and
on the fermions are given by 
\begin{align}
\label{veff_b}
  V_{eff}^b(r)&=V_{ext}^b(r)+2g_{bb}n_c(r)+2g_{bb}n_{nc}(r)+g_{bf}
 n_f(r)
\end{align}
{\rm and}  
\begin{align}
\label{veff_f}
V_{eff}^f(r)&=V_{ext}^f(r)+g_{bf}n_c(r)+g_{bf}n_{nc}(r),
\end{align}
where we have introduced the notations $\beta=(K_BT)^{-1}$, $g_{bb}=4 \pi
\hbar^2 a_{bb}/m_b$ and $g_{bf}=2 \pi  
\hbar^2 a_{bf}/m_r$ with
$m_r^{-1}=m_b^{-1}+m_f^{-1}$. The chemical potentials $\mu_b$ and $\mu_f$ are
determined by the normalization conditions
$ N_b=\int \left( n_c(r)+n_{nc}(r) \right)\, d^3r$ and
$N_f=\int  n_f(r)\, d^3r$,
which ensure the self-consistent closure of the model. 

Equations~(\ref{nc}-\ref{veff_f}) have been derived from a grand-canonical
Hamiltonian in 
which  the interactions are included in a mean-field Hartree-Fock
approximation \cite{thermodBEC}, by  employing 
the  semiclassical approximation for the bosonic thermal cloud and for
the fermions and by taking the strong coupling limit
$N_b a_{bb}/a_{ho} \gg1$ for the wave-function of the condensate,
with $a_{ho}=(\hbar/m_b \omega_b)^{1/2}$ 
the bosonic harmonic oscillator length.
Upon averaging the Hamiltonian
on the equilibrium state of the system at finite temperature we obtain 
the energy as
the sum of various contributions: the kinetic and the external
confinement energy for each of the species, as well as the boson-boson
and 
boson-fermion interaction terms. One has
\begin{multline} \label{en:finT}
E=
E_{kin}^f+E_{ext}^f+E_{kin}^b+E_{ext}^b+E_{int}^{bb}+E_{int}^{bf}\\=\frac
3 2 \left( \frac{m_f}{2 \pi \hbar^2} \right)^{3/2} \beta^{-5/2} 
\int f_{5/2}(z_f)\, d^3r+
\frac 3 2 \left( \frac{m_b}{2 \pi \hbar^2} \right)^{3/2} \beta^{-5/2}
\int g_{5/2}(z_b)\, d^3r+\\
\int V^b_{ext}(r) \left( n_c(r)+n_{nc}(r) \right)\, d^3r+
\int V^f_{ext}(r)n_{f}(r)\, d^3r+ \\ \frac {g_{bb}} 2 \int
\left(n_c^2(r)+2 n_{nc}^2(r)+4n_c(r)n_{nc}(r) \right)\, d^3r+
 g_{bf} \int \left( n_c(r)+n_{nc}(r) \right) n_f(r)\, d^3r
\end{multline}
where 
$f_p(z)=\Gamma(p)^{-1} \int y^{p-1}\, dy/
(z^{-1}e^y+1)$, $g_p(z)=\Gamma(p)^{-1} \int y^{p-1}\, dy/
(z^{-1}e^y-1)$ are the usual Fermi and Bose functions and
$z_{f,b}=\exp(\beta(\mu_{f,b}-V_{eff}^{f,b}(r)))$ .
The release energy is obtained by setting the confinement potentials
$V_{ext}^{b,f}$ in eq.~(\ref{en:finT}) to zero. 
This quantity can be
measured via time-of-flight experiments.

At low temperature the thermal component $n_{nc}$ can be safely neglected
in the right hand side of equations (\ref{nc}),(\ref{veff_b}) and
(\ref{veff_f}).
 Similarly,
when the fermionic density $n_f$ is much
smaller than the density $n_c$ of the Bose condensate, its contribution
in the right hand side of the same equations can be dropped. 
This is valid, for example, if the interaction strengths $g_{bf}$ and
$g_{bb}$ have comparable size and if the 
trapping potential $V_{ext}^f$ is not too stiff with respect to
$V_{ext}^b$. In this case the density profile of the condensate is
not affected by the interaction with the fermions and the effective potential
felt by the fermions takes the simplified form:
\begin{equation} \label{V:simpmod}
V^f_{eff}(r)=
\begin{cases} \displaystyle
\frac 1 2 m_f {\omega}_f^2(1-\gamma) r^2
+ \frac {g_{bf}} {g_{bb}} \mu_b & \text{for $r < R_b$}\\ \displaystyle
\frac 1 2 m_f \omega_f^2 r^2 & \text{for $r \ge R_b$}
\end{cases}
\end{equation}
where
\begin{equation} \label{gamma}
\gamma=\frac{g_{bf}}{g_{bb}} \frac{ m_b \omega_b^2} {m_f \omega_f^2}
\end{equation}
and $R_b=(2\mu_b/m_b\omega_b^2)^{1/2}$ is the radius of the condensate
cloud. 
The potential (\ref{V:simpmod}) depends on temperature
through the boson chemical potential $\mu_b$, which determines the
radius $R_b$. Full numerical calculations, including the contribution of the thermal
Bose and Fermi components, show that this simplified model (hereafter
called the double-parabola model)
describes very well the main features of the system below
the critical temperature for Bose-Einstein condensation.

In the double-parabola model, if the parameter  $\gamma$ in eq.
(\ref{gamma}) is smaller than unity, the potential felt by the
fermions has its  minimum in the center of the trap. In this situation two
limiting cases can be envisaged by  comparing $R_b$ with the radius
of the Fermi cloud. This is approximatively  given by the Fermi radius
$R_F=(2E_F/m_f \omega_f^2)^{1/2}$ calculated in the absence of
interactions, where $E_F=(6N_f)^{1/3}\hbar \omega_f$ is the Fermi
energy.
In the limit $R_b \ll R_F$ the 
number of bosons is much smaller than the number of fermions and thus the
interactions play a minor role. Instead, in the limit $R_b \gg
R_F$ the fermionic cloud feels a harmonic trapping potential
with a renormalized frequency $\tilde \omega_f =
\omega_f (1-\gamma)^{1/2}$.
Finally in the case $\gamma >1$ the repulsive interaction with the bosons is
stronger than the external potential. The effective potential
(\ref{V:simpmod}) then
exhibits a
local maximum at the center of the trap.

\section{Scaling and role of the interactions at $T=0$}

\label{sec2}

Let us begin our discussion in the framework of the double-parabola
model introduced in the previous section.
In the case $\gamma< 1$ we can give
a simple approximate solution of the model by a variational minimization
of the energy functional in eq.~(\ref{en:finT}). At $T=0$ this  reads
\begin{multline} \label{en:T=0}
E(T=0)= \frac 3 5 \frac{\hbar^2}{2m_f} (6 \pi^2)^{2/3} \int
n_f^{5/3}(r)\, d^3r + \int V^b_{ext}(r) n_c(r)\, d^3r+\\
\int V^f_{ext}(r) n_f(r)\, d^3r+ \frac {g_{bb}} 2 \int n_c^2(r)\, d^3r
+g_{bf} \int n_c(r) n_f(r) \, d^3r.
\end{multline}
The variational approach (see the details in the Appendix)
explicitly shows that the relevant properties of the system
depend on the various parameters of the model through two adimensional
combinations, which are the parameter $\gamma$ in  eq.~(\ref{gamma})
and
a parameter $x$ given by
\begin{equation}
\label{x}
x=\sqrt{\frac{R_b}{R_F}}=\sqrt{\frac{m_f\omega_f}{2m_b\omega_b}} 
\left(\frac{15a_{bb}}{a_{ho}} \frac{N_b}{(6N_f)^{5/6}}\right)^{1/5}.
\end{equation} 
At given $\gamma$, the  ratio of the sizes of
the two clouds in the absence of interactions determines the deviation
of the kinetic energy of the Fermi component from its ideal-gas value.

We have checked numerically that the scaling in these two variables is
satisfied with good accuracy also by the full 
numerical solution of  eqs.~(\ref{nc}-\ref{veff_f}) at zero
temperature for any value of $\gamma$. 
The description of the  system with
only two scaling parameters instead of the eight original ones
entering eqs.~(\ref{nc}-\ref{nf}) represents a major simplification of
the problem.
In view of this property, in the following we shall
present our discussion in terms of the scaling parameters $x$ and
$\gamma$.

\begin{figure}
\centerline{\epsfig{figure=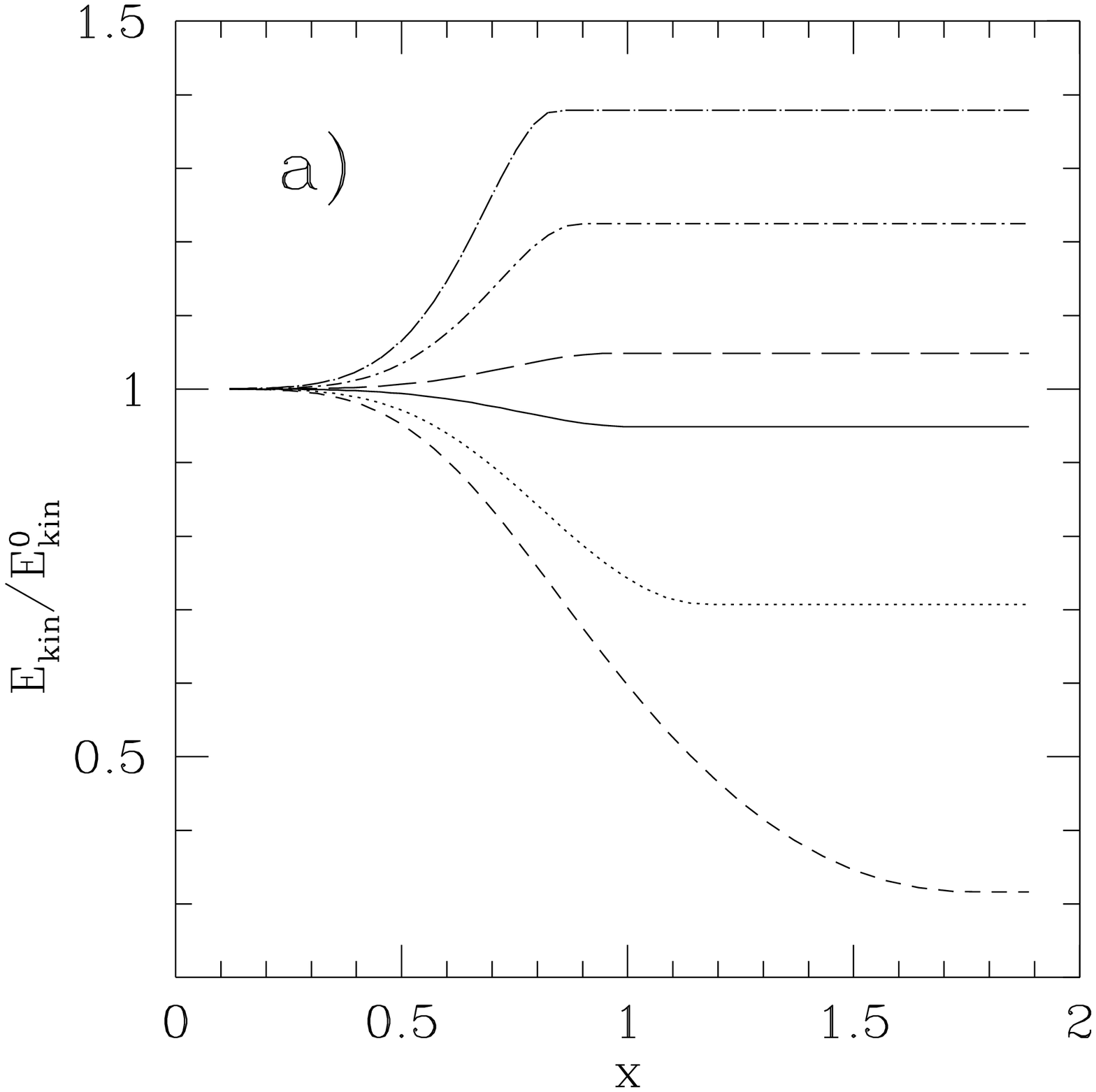,width=0.5\textwidth}
\epsfig{figure=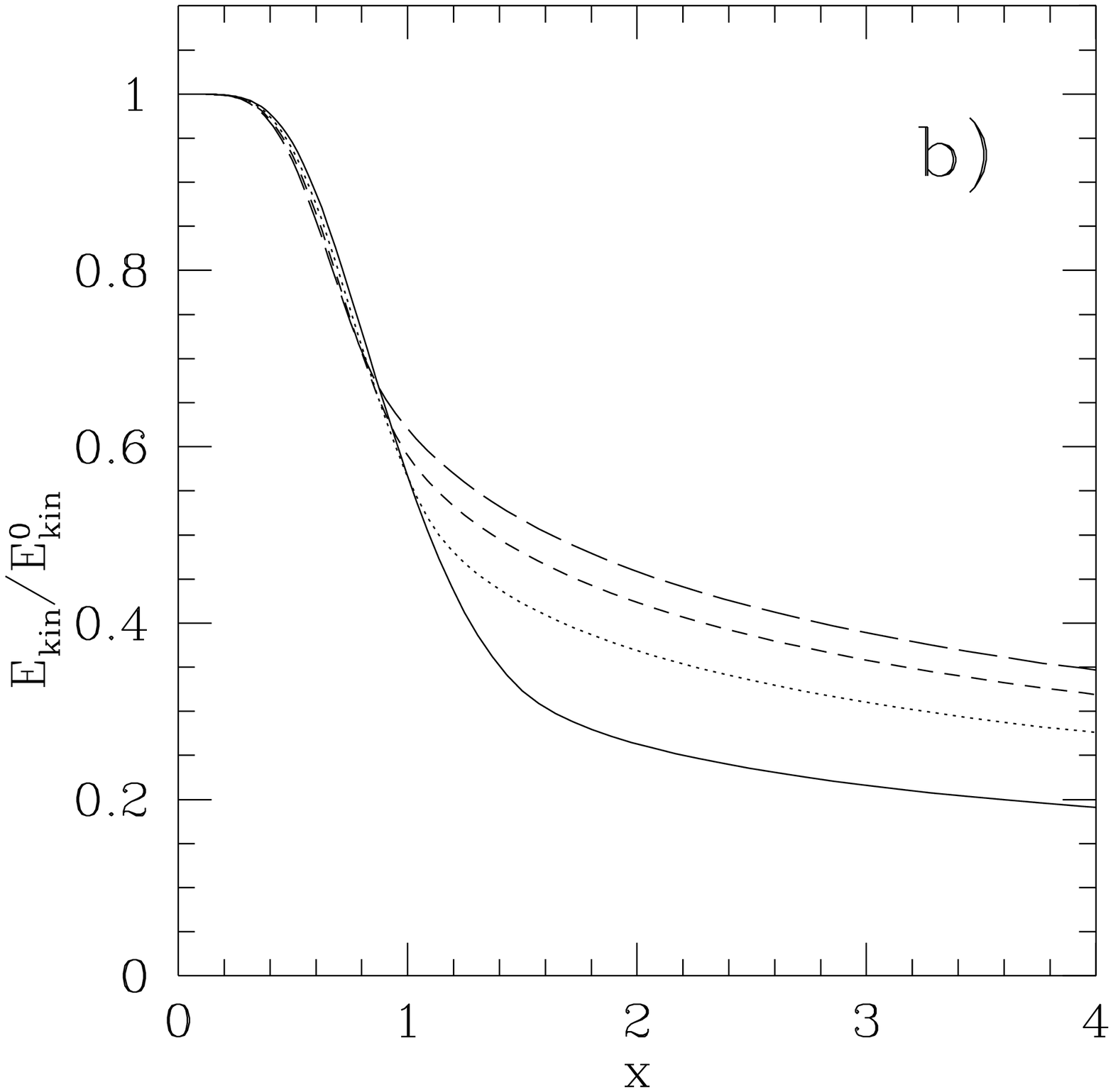,width=0.5\textwidth}}
\caption{Kinetic energy of the Fermi component in units of $E_{kin}^0=
 3 N_f E_F/8$ as a function of the
scaling parameter $x$ for different  values of $\gamma$. a): for $\gamma <
1 $ taking the following values: $\gamma=0.9$ 
(short dahses), $\gamma=0.5$ (dots), $\gamma=0.1$ (solid),
$\gamma=-0.1$ (long dashes), $\gamma=-0.5$ (dot-short dashes) and
$\gamma=-0.9$ (dot-long dashes). b):
for $\gamma >1$ ranging from $\gamma=1.1$ (solid curve) to
$\gamma=1.7$ (long dashed curve) in steps of 0.2.}
\label{fig2}
\end{figure}

In fig.~\ref{fig2}.a we show a plot of the kinetic energy as a
function of $x$ at zero temperature for
different values of $\gamma< 1$. As $x$ increases,
the kinetic energy of the fermions goes from its
non-interacting value $E^0_{kin}=3N_fE_F/8$ to the
strong-coupling limit $\tilde E_{kin}=E^0_{kin}(1-\gamma)^{1/2}$.
As a first result of our analysis we see from fig.~\ref{fig2}.a
that there is a clear correspondence,
for a fixed value of $x$, between the value of the 
kinetic energy and the value of $\gamma$. Therefore the sign and the
strength of the ratio between the boson-fermion and boson-boson 
coupling constants could be inferred from a measurement of the fermion
kinetic energy.

In the case $\gamma > 1$
the fermions are expelled from the center of the trap \cite{PiTn1,Molmer2}
and form a
shell around 
the bosons as $N_b$ increases with respect to $N_f$. In this case the
kinetic 
energy of the fermions
(fig.~\ref{fig2}.b) tends to
zero when $x \to \infty$.

\begin{figure}
\centerline{\epsfig{figure=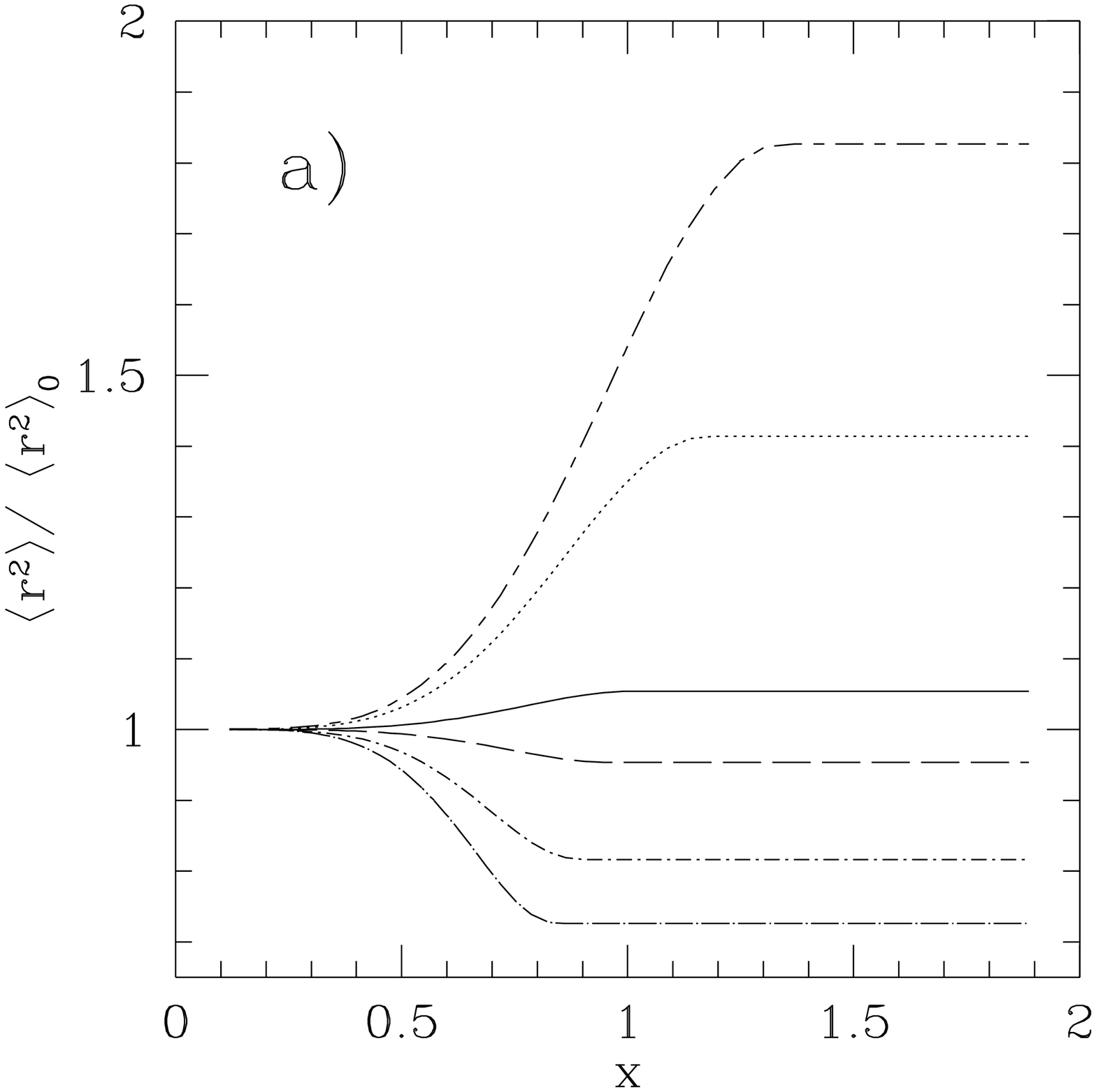,width=0.5\textwidth}
\epsfig{figure=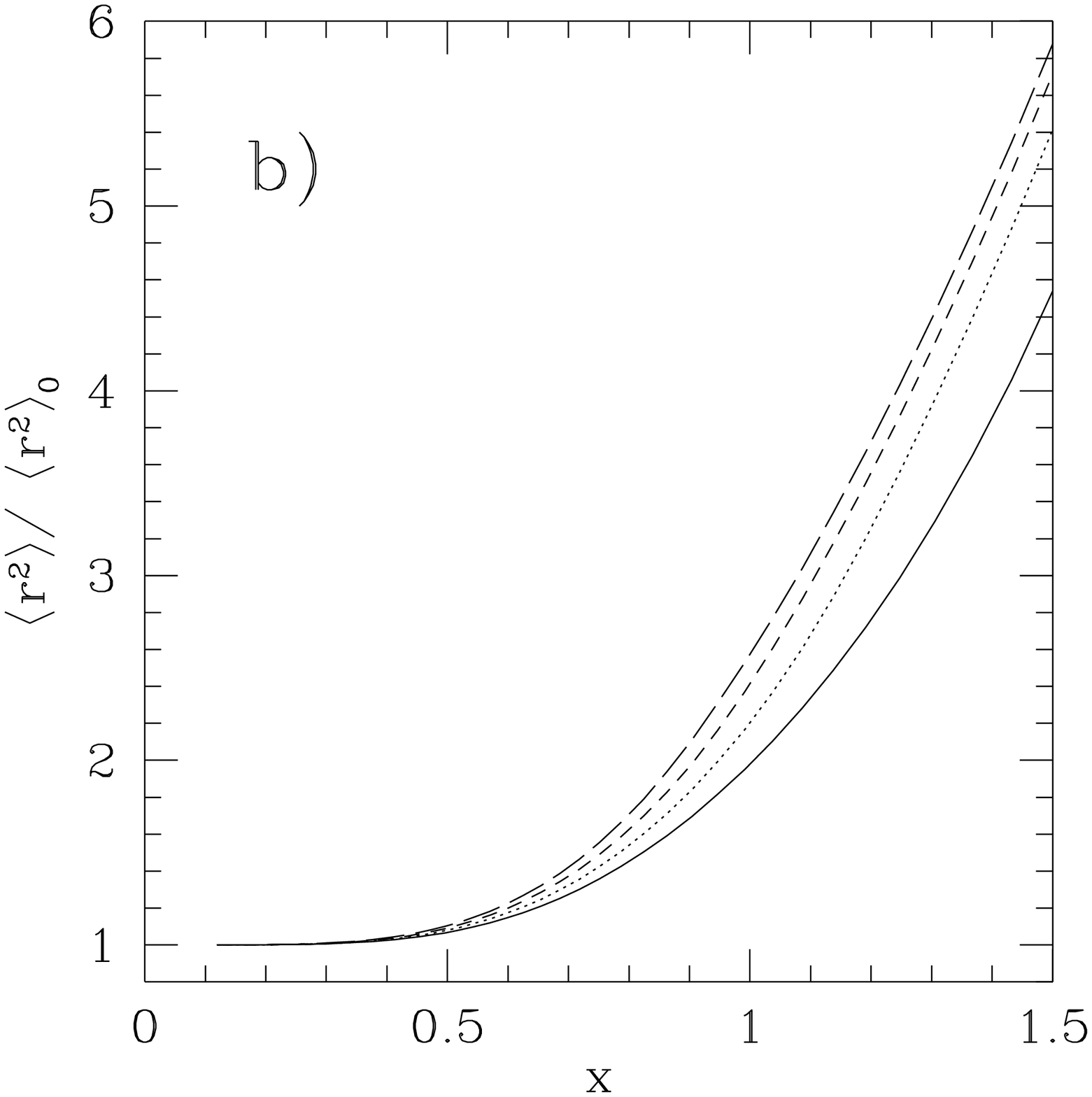,width=0.5\textwidth}}
\caption{Mean square radius of the Fermi component in units of
$\langle r^2 \rangle_0 = 3 N_f R_F^2/8$
as a function of the
scaling parameter $x$ for different  values of $\gamma$. a): same
values and notations as in figure~\ref{fig2}.a except for the value
$\gamma=0.7$ (long dashes--short dashes) which replaces the value
$\gamma=0.9$ of figure~\ref{fig2}.a.
 b): for 
$\gamma >1$, ranging from $\gamma=1.1$ (solid curve) to
$\gamma=1.7$ (long dashed curve) in steps of 0.2.}
\label{fig3}
\end{figure}

For completeness we have also analyzed the behaviour of the mean
square radius of the fermionic cloud 
as a function of $x$.
For $\gamma
 < 1$ the asymptotic value at large
 $x$ is larger(smaller) than in the ideal case for
$\gamma > 0(<0)$ (see fig.~\ref{fig3}.a).
For $\gamma > 1$ the mean square radius increases indefinitely with
increasing $x$ 
(fig.~\ref{fig3}.b). These behaviours are immediately understood in
terms of the behaviour of the kinetic energy shown in fig.~\ref{fig2}.

\section{The role of temperature}

\begin{figure}
\centerline{
\epsfig{figure=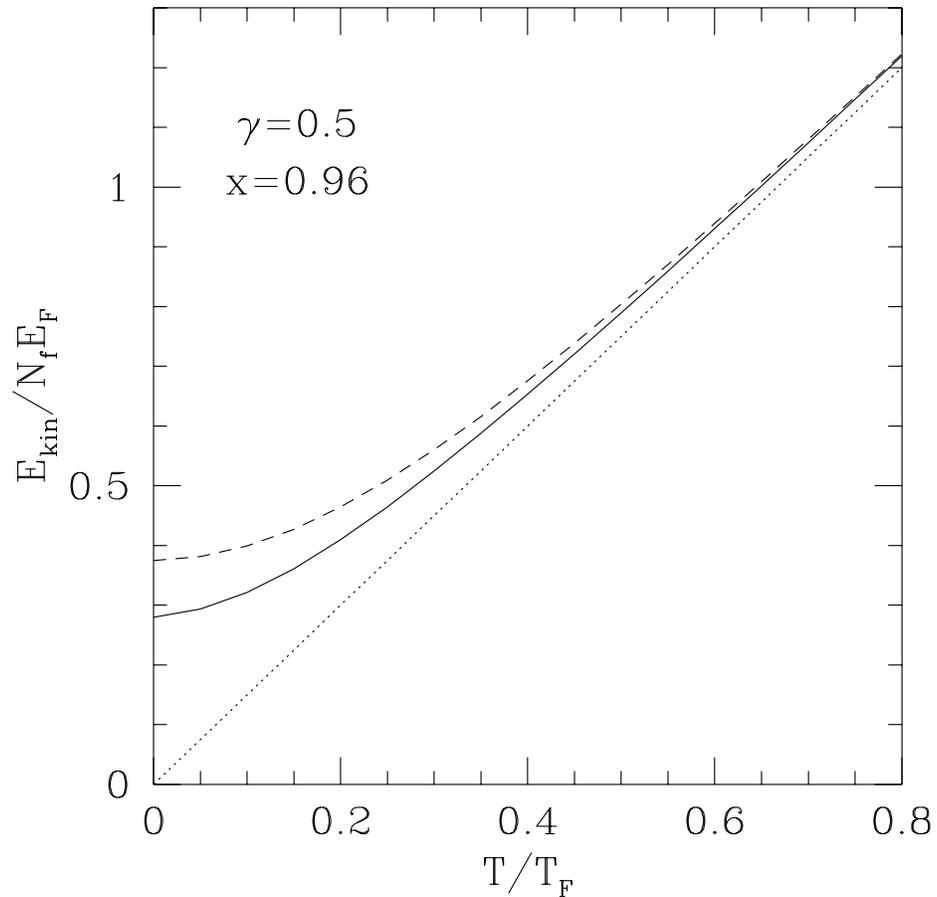,width=0.7\textwidth}}
\caption{Kinetic energy of the Fermi component as
a function of the reduced temperature $T/T_F$ for a choice of
the scaling parameters corresponding to $m_f=m_b=39\ a.u.$,
$\omega_f=\omega_b=2 \pi 
\times 100 \ sec^{-1}$, $N_f=10^4$, $N_b=10^6$, $a_{bb}=92\,a_0$ and
$a_{bf}=46\, a_0$ where $a_0$ is the Bohr radius. 
The solid curve refers to the interacting
system and  the dashed line to the non-interacting one. The dotted curve
shows the classical result.}
\label{fig4}
\end{figure}

\begin{figure}
\centerline{\epsfig{figure=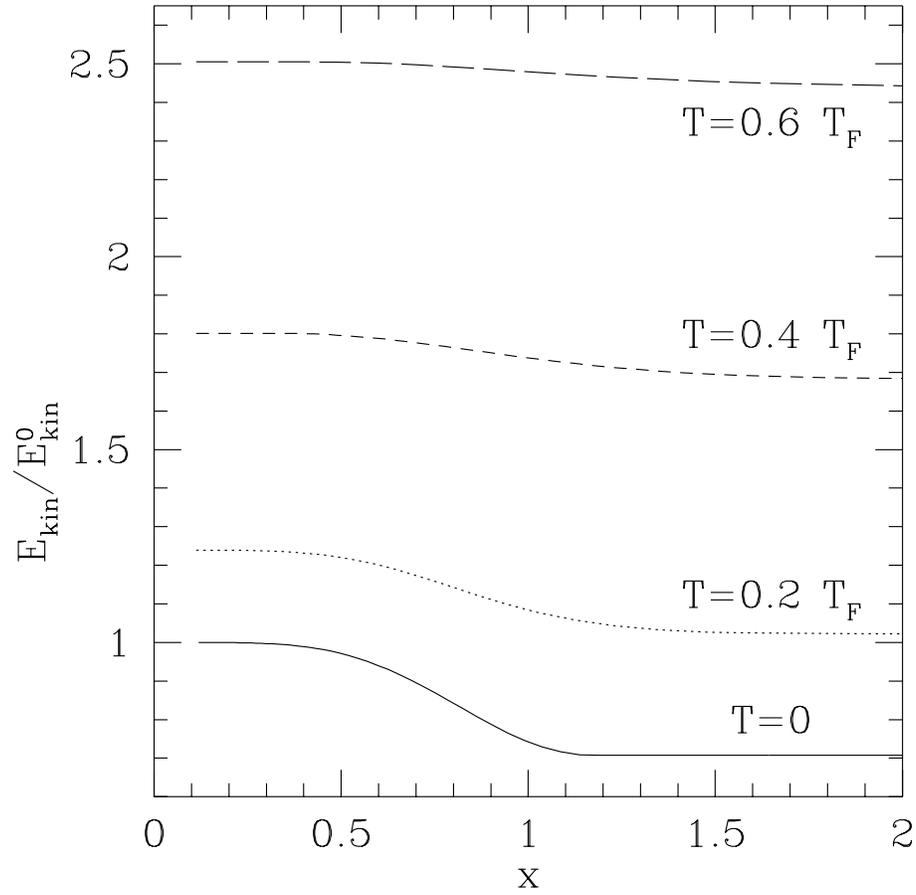,width=0.7\textwidth}}
\caption{Kinetic energy of the Fermi component as a function of the
scaling parameter $x$ for $\gamma=0.5$ at different temperatures, as
indicated in the figure.}
\label{fig5}
\end{figure}

Let us finally examine the temperature
dependence of the kinetic energy of the Fermi gas.
To this purpose we have solved self-consistently the
full set of eqs.~(\ref{nc}-\ref{veff_f}). 
Of course, in the classical
regime the kinetic energy is insensitive to
interactions.
As quantum
degeneracy sets in at $T<T_F$, where $T_F=E_F/K_B$ is the Fermi temperature,
deviations 
from the classical value $3N_fK_BT/2$
become apparent. 
We show in fig.~\ref{fig4} the predicted behaviour for a given choice
of the parameters of the mixture. The role of the interactions
decreases as temperature increases. This
can also be seen in fig.~\ref{fig5} where we plot the kinetic energy
as a function of $x$ for a choice of different temperatures.

The scaling behaviour described in sec. \ref{sec2} 
is less
accurate at finite temperature. This is easily understood from the
fact that
the approximations leading to
eq.~(\ref{V:simpmod}) become less justified as temperature increases.

\section{Conclusions}
\label{sec4}
We have presented a broad study of the kinetic energy of the
fermionic component in a mixture of bosons and fermions in the so-called
Thomas-Fermi regime ($N_b a_{bb}/a_{ho} \gg 1$).
We have shown that at zero temperature the kinetic energy, as well as
the mean square radius of the Fermi component, exhibit an important
scaling behaviour in the relevant parameters $\gamma$  and $x$ defined
in eqs.~(\ref{gamma},\ref{x}). This has allowed us to give a
systematic 
investigation of
these physical properties on a wide range of system parameters 
 and  to understand the role of
the interactions between the Fermi gas and the Bose condensate.

In particular, we have found that the shift of the fermionic kinetic energy due
to the interactions becomes sizeable  at appreciable values of the
parameter $x$ mesauring the relative radii of the two clouds. This
effect could be used to infer the sign of the boson-fermion scattering
length from measurements of the fermion kinetic energy.

Finally, the role of temperature has been investigated within the
self-consistent numerical solution of the full set of equations for the
coupled boson-fermion mixture and the interplay
between thermal and interaction effects on the kinetic energy has been
demonstrated.

This work is supported by the Istituto Nazionale per la Fisica della Materia
through the Advanced Research Project on BEC. One of us (L.V.)
acknowledges the hospitality of the Scuola Normale Superiore di Pisa
during part of this work. 

\appendix
\section{Variational model}
\label{sec3}

The scaling behaviour discussed in section \ref{sec2} can be
explicitly predicted by a
variational approach which turns out to be very accurate in
reproducing the
numerical results at $T=0$ in the case $\gamma<1$.
The variational method is based on the minimization of the energy
functional 
 (\ref{en:T=0})
within a restricted class of functions. 

For $\gamma < 1$ it is convenient to describe the
fermionic cloud as if it were
embedded in an effective potential $V_{var}(r)=\frac 1 2 m_f
\omega_{var}^2 r^2$, where the frequency $\omega_{var}$ is a
variational parameter. The corresponding fermionic density profile is
$n_f(r)=1/ 
(6\pi^2) (2m_f/\hbar^2 )^{3/2}( E_F^{var} - m_f \omega_{var}^2 r^2/2)^{3/2}$
and the  expression
for the variational energy functional takes the form
\begin{multline} \label{En:ferm}
E_{var}=E_{kin}+E_{ho}+E_{int}=
\frac 3 8 N_f E_F \frac{\omega_{var}}{\omega_f} + \frac 3 8 N_f E_F
\frac{\omega_f}{\omega_{var}}+\\ \frac {g_{bf}} {g_{bb}}
\int^{\min(R_F^{var},R_b)}
\frac 1 {6\pi^2} \left(\frac{2m_f}{\hbar^2} \right)^{3/2}
\left( E_F^{var} -\frac 1 2 m_f \omega_{var}^2 r^2
\right)^{3/2} \left(\mu_b -\frac 1 2 m_b\omega_b^2 r^2 \right)\, d^3r.
\end{multline}
Here $E_F^{var}=(6N_f)^{1/3}\hbar \omega_{var}$
and $R_F^{var}=(2E_F^{var}/m_f\omega_{var}^2)^{1/2}$ are the the Fermi
energy and the 
Fermi radius calculated with the frequency $\omega_{var}$.
The bosons are described by the Thomas-Fermi inverted parabola
$n_b(r)=g_{bb}^{-1}(\mu_b-m_b \omega_b^2 r^2/2)$ with $\mu_b=\hbar \omega_b
(15N_b a_{bb}/a_{ho})^{2/5}/2$.

The integral in eq.~(\ref{En:ferm})can be carried out analytically, 
with the result
\begin{equation} \label{e_var}
E_{var}(x,\gamma,\alpha)=\frac 38 N_f E_F \times 
\begin{cases} \displaystyle \frac{\alpha^2}{x^2}+\frac{x^2}{\alpha^2}+
 \gamma \frac{x^2}{\alpha^2} P(\alpha) & 
\text{for $\alpha< 1$} \\ \displaystyle
\frac{\alpha^2}{x^2}+\frac{x^2}{\alpha^2}+ \gamma \frac{x^2}{\alpha^2}
\left( -1+\frac 8 3 \alpha^2 \right) &
\text{for $\alpha \ge 1$}
\end{cases}
\end{equation}
where $\alpha^2=x^2\omega_{var}/\omega_f$ and
\begin{equation}
P(\alpha)=
\frac 2 {9 \pi} \left[ \alpha \sqrt{1-\alpha^2}( 9 -18 \alpha^2 
 + 40 \alpha^4- 16 \alpha^6  ) +3 (-3+8 \alpha^2)
\arcsin (\alpha) \right]  ,
\end{equation}
\begin{equation}
x=\sqrt{\frac{m_f \omega_f}{2 m_b \omega_b}}\left( 15 \frac{a_{bb}}{a_{ho}}
\frac{N_b}{(6N_f)^{5/6}}
\right)^{1/5}.
\end{equation}

The condition $\partial E_{var}/\partial \alpha =0$ determines the
value of $\alpha$ and hence of $\omega_{var}$.
This equation has to be solved numerically,  except for
 $\alpha \ge 1$ where
the model gives the result $\omega_{var}=\tilde{\omega}_f$.

The expression (\ref{e_var}) allows an explicit identification
of the scaling variables introduced in sec.~\ref{sec2}.
In fact, the quantity $E_{var}/N_fE_F$ at its minimum depends only on $x$ and
$\gamma$.

Of course the variational estimate  gives an upper bound for the total
energy. This bound is very close to the value obtained by solving
the Schr\"odinger equation with the potential
(\ref{V:simpmod}). Typical deviations are less than 1 \% of the energy.


\end{document}